\documentclass{scrartcl}
\usepackage{amsmath,amssymb,units,graphicx,color}
\usepackage{natbib}

\setkomafont{caption}{\normalfont\normalcolor\small}
\setkomafont{captionlabel}{\normalfont\normalcolor\sffamily\bfseries\small}
\setcapindent{0em}

\newcommand*{\stern}      {\ensuremath{\star}}

\newcommand*{\Msun}       {\ensuremath{\text{M}_{\odot}}}

\newcommand*{\rhostar}    {\ensuremath{\rho_{\stern}}}

\newcommand*{\SNu}        {\ensuremath{\operatorname{SNu}}}

\title{Evolution of the ISM and Galactic Activity} 
\author{Volker Gaibler\footnote{Contact: \texttt{V.Gaibler@lsw.uni-heidelberg.de}} , 
Max Camenzind, Martin Krause\\Landessternwarte Heidelberg--K\"onigstuhl}
\date{}

\begin{document}
\maketitle


\section{Introduction}
We examine the impact of time-dependent mass injection and heating on the
evolution of the interstellar medium (ISM) in elliptical galaxies.  As the large
and luminous ellipticals have supermassive black holes at their cores, which
were probably much less massive in the young universe, feeding these black holes
is essential. The evolution of the ISM, controlled by supernova and/or AGN
heating as well as cooling by bremsstrahlung \cite{Mathews2003}, could provide
an abundant matter supply for this if a sufficient fraction of the ISM flows
towards the center.

The hot phase of the ISM has typical temperatures around $10^{7}\,\mathrm{K}$
at the galactic cores. While stellar winds and planetary nebulae are widely
seen as the sources of the interstellar gas, there is no generally established
model to explain the heating of the ISM (supernovae, AGN heating, collisions).
We concentrate on supernovae as a source of the heating.

\section{Model and Basic Equations}
We use simple Plummer models as well as combined Hernquist/NFW models for the mass
distribution of our galaxies, with $\rho_\star(r)$ being the stellar mass
density. In addition, we have a black hole with mass $M_{\mathrm{H}}$ at the
center. Most calculations are performed in one spatial dimension, some in 2D
including random disturbances, with the magneto-hydrodynamic code NIRVANA by Udo
Ziegler \cite{Ziegler1997}.

\begin{align}
  \frac{\partial \rho}{\partial t} + \nabla \cdot(\rho {\mathbf v}) &= \alpha
  \rhostar
  \label{eqn:kontigleichung}
  \\
  \frac{\partial (\rho {\mathbf v})}{\partial t} + \nabla \cdot
  (\rho\mathbf{v}\otimes\mathbf{v}) &= - \nabla p - \rho \nabla\Phi
  \label{eqn:impulsgleichung}
  \\
  \frac{\partial e}{\partial t} + \nabla \cdot(e {\mathbf v})&= -p \nabla
  \cdot {\mathbf v} 
  +\alpha \rho_* c_{\text v} T_0 
  -\rho^2 \Lambda 
  \label{eqn:energiegleichung}
\end{align}

Here $\rho$, ${\bf u}$, $T$, $p$ and $e$ (energy density) are the state
variables of the gas, $\rho^2 \Lambda$ is the cooling rate, and $c_{\mathrm v}
T_{0}$ is the supplied energy relative to the injected matter $\alpha
\rho_\star$, so the heating is controlled by the value of $T_0$. The mass
injection by stellar mass loss as well as the supernova heating are assumed to
be proportional to the stellar density because of their physical origins. 
\begin{figure}
  \centering
  \includegraphics[width=9cm]{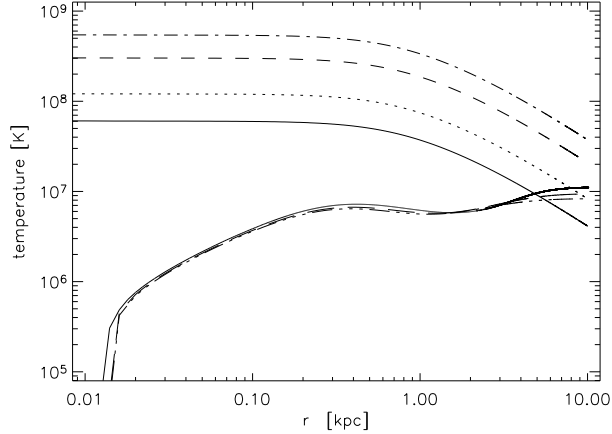}
  \caption{Typical temperature profiles of steady state outflow and inflow solutions.
  The two types of solutions can be easily distinguished in this figure: While
  inflows generally have low temperatures and positive gradients, galactic winds
  show high temperatures and negative gradients, which might be very difficult to
  observe directly as the densities are much lower.
  $\alpha=\unit[10^{-19}]{s^{-1}}$, inflows: $T_0=\unit[5\cdot
  10^5]{K}\ldots\unit[10^7]{K}$, outflows: $T_0=\unit[1\cdot
  10^8]{K}\ldots\unit[9\cdot 10^8]{K}$.}
\end{figure}

\section{ISM Simulations}
For constant mass injection and heating we find steady state solutions, which
can be characterized as inflow and outflow types (Figure 1). Inflow solutions
show temperatures near the virial temperature and subsonic gas flows towards the
center (cooling flows) independent of the heating rate. Outflows on the other
hand generally are supersonic, of high temperature and low gas density. Their
velocity and temperature profile is independent of the mass injection rate. 
\begin{figure}
  \centering
  \includegraphics[width=9cm]{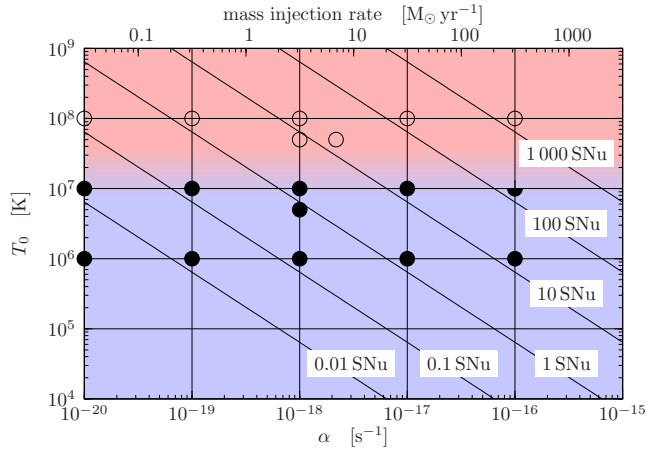}
  \caption{Distribution of inflow (filled symbols) and outflow (open symbols)
  solutions in the $(\alpha, T_0)$ parameter space for our model galaxy (dark
  matter contribution is not included). The
  mass injection parameter $\alpha$ directly corresponds to the overall mass
  injection rate in the galaxy, while the heating parameter $T_0$ depends on the
  supernova rate (in $\SNu$) and the mass injection. $\unit[1]{\SNu}$
  corresponds to 1 supernova per 100 yr and $\unit[10^{10}]{\mathrm{L}_\odot}$. 
  The diagonal lines represent constant supernova rates.}
\end{figure}

Although there is a transition region between inflow and outflow in the
$(\alpha, T_0)$ parameter space where we have spatially bimodal flows, this
region seems to be narrow. If there exists at least a partial inflow, then a
significant fraction of the globally injected mass flows towards the center. 
\begin{figure}
  \centering
  \includegraphics[width=9cm]{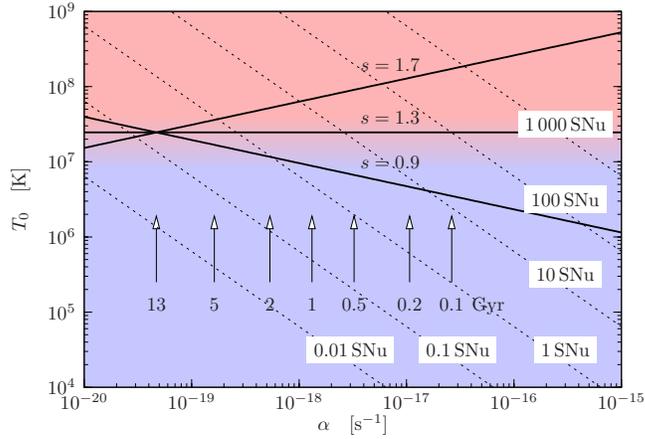}
  \caption{Evolution of SN Ia heating in the $(\alpha, T_0)$ parameter space for
  a powerlaw behavior at late times in the galaxy: 
  $\text{supernova rate} \propto t^{-s}$ and mass injection $\alpha \propto t^{-1.3}$.
  The arrows indicate the $\alpha$-position of the galaxy at a given time.}
\end{figure}

Figure 2 shows the distribution of the two solution types in the parameter
space. There obviously exists a critical heating parameter $T_{0, \text{crit}}$
which separates these types. This can be explained easily by checking the
energy balance. The ISM evolution in our simulations with time-dependent mass
injection and heating can be understood by examining the evolution of the model
in the $(\alpha, T_0)$ parameter space. 

The time-dependence of mass injection follows from the initial mass function of
the stellar population generated by a monolithic starburst. The amount of mass
injection seems to be in good agreement with observations \cite{Bregman2004}. 

The time-dependence of the supernova rate is rather unknown. During the initial
starburst there is much contribution from SNe II, but soon after SNe Ia are the only
ones occurring. At late times, in passively evolving galaxies, the evolution of 
the SN Ia heating rate is often assumed to be a powerlaw ($\propto t^{-s}$), 
but observational confirmation for this is still missing so that the powerlaw
exponent could vary over a large range. Figure~3 shows the evolutionary tracks
of the model for three different powerlaw exponents. 

The critical heating parameter is governed by the virial mass of the galaxy, and
supernova rates in local ellipticals \cite{Cappellaro1999} suggest that our
model galaxy of $10^{11} \Msun$ virial mass should be near the turnover region
between inflow and outflow. For early times, we find a galactic wind for $s=1.7$
and inflow for $s=0.9$. This shows that higher supernova rates in the past
($s>0$) are not sufficient to drive galactic winds --- they have to increase
stronger ($s>1.3$) than the mass injection \cite{Ciotti1991}. 

For a galaxy with much higher or much lower mass, the position of the turnover
region is much more important than the evolution of the supernova rate so that
high-mass galaxies would generate inflows with high X-ray luminosities while 
low-mass dwarf galaxies would have strong galactic winds and unobservable
X-ray luminosities.

\section{Impact of the Initial Starburst}

At very early times in the galaxy, during its formation, the abovementioned
powerlaw behavior of course breaks down.  Figure 4 shows the evolution of the
parameters $\alpha$ and $T_0$ during galaxy formation based on two very simple
scenarios with data from Starburst99 \cite{Leitherer1999}: The galaxy is
\emph{instantaneously} formed from a gas cloud or the galaxy is formed during
500 Myr by a \emph{constant star formation rate}. The mass injection $\alpha$
decreases with increasing formation timescale, but the heating parameter $T_0$
remains the same because $T_0$ is proportional to the heating rate per injected
mass.

The extremely high supernova rates during the starburst lead to high heating
rates, but this does not necessarily lead to galactic winds. Figure 4 shows
that $T_0\sim \unit[10^8]{K}$ during the starburst. This would drive a galactic
wind for the model galaxy, but couldn't do so for a galaxy with a virial mass of
$\gtrsim\unit[10^{12}]{\Msun}$ as the threshold-$T_0$ for winds increases
proportional to the galaxy mass. The reason is that a huge amount of injected
matter would have to be lifted out of the galaxy's gravitational potential.

\section{Effects on Black Hole Growth}
The evolution of the ISM might be a crucial factor for fuelling central black
holes. If a galactic wind arises and stays for a long time, this could
effectively avoid black hole growth because of low gas densities and high
temperatures. A big galaxy in inflow state, on the other hand, could easily
accrete some $\unit[10^{10}]{\Msun}$ because the total injected gas mass is of
the order of the stellar mass of the galaxy. 

Depending on the galaxy mass and the evolution of the heating rate, there is a
rich variety of possible scenarios for the interstellar gas. Inflow at early
times can produce supermassive black holes very quickly and possibly resulting
activity could provide a feedback through additional heating which could turn
the inflow into an outflow. The impact of further star formation in cooling
flows could be a similar feedback mechanism and induce a cyclic behavior.
\begin{figure}
  \centering
  \includegraphics[width=9cm]{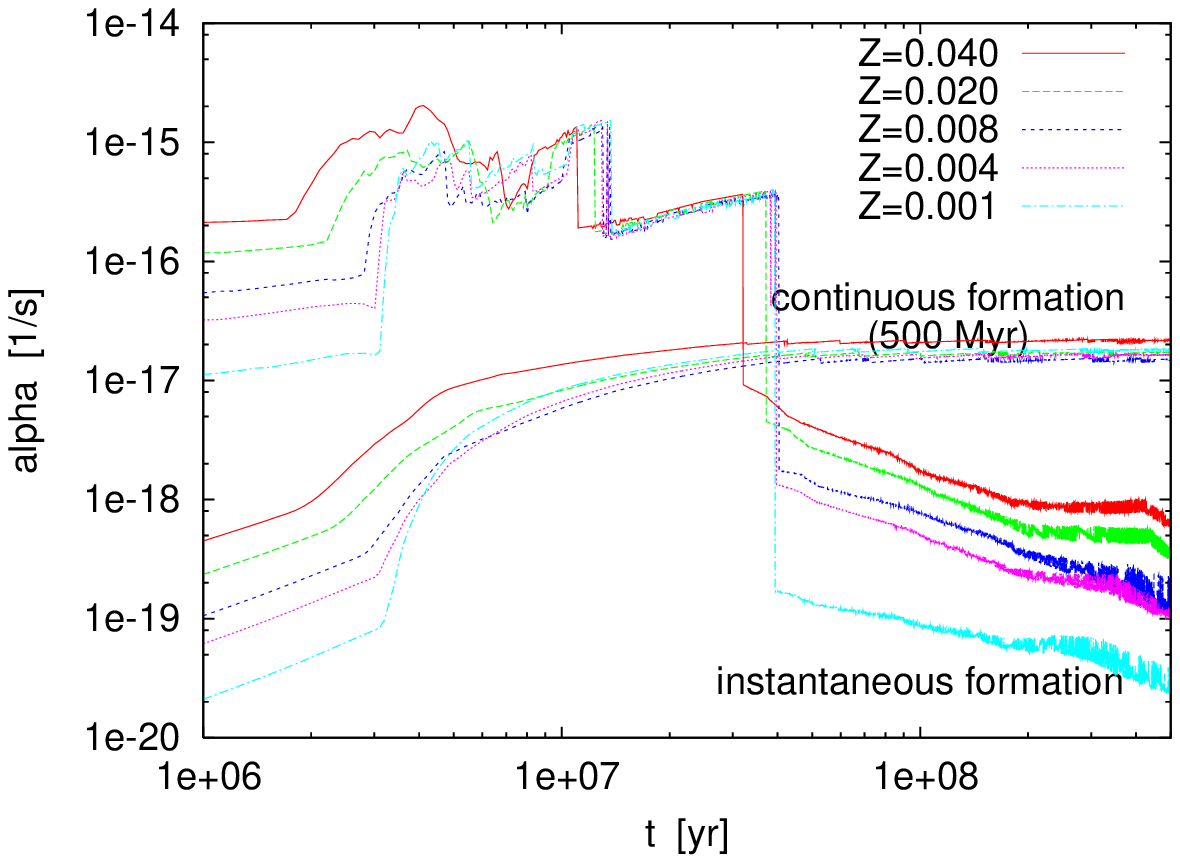}
  \includegraphics[width=9cm]{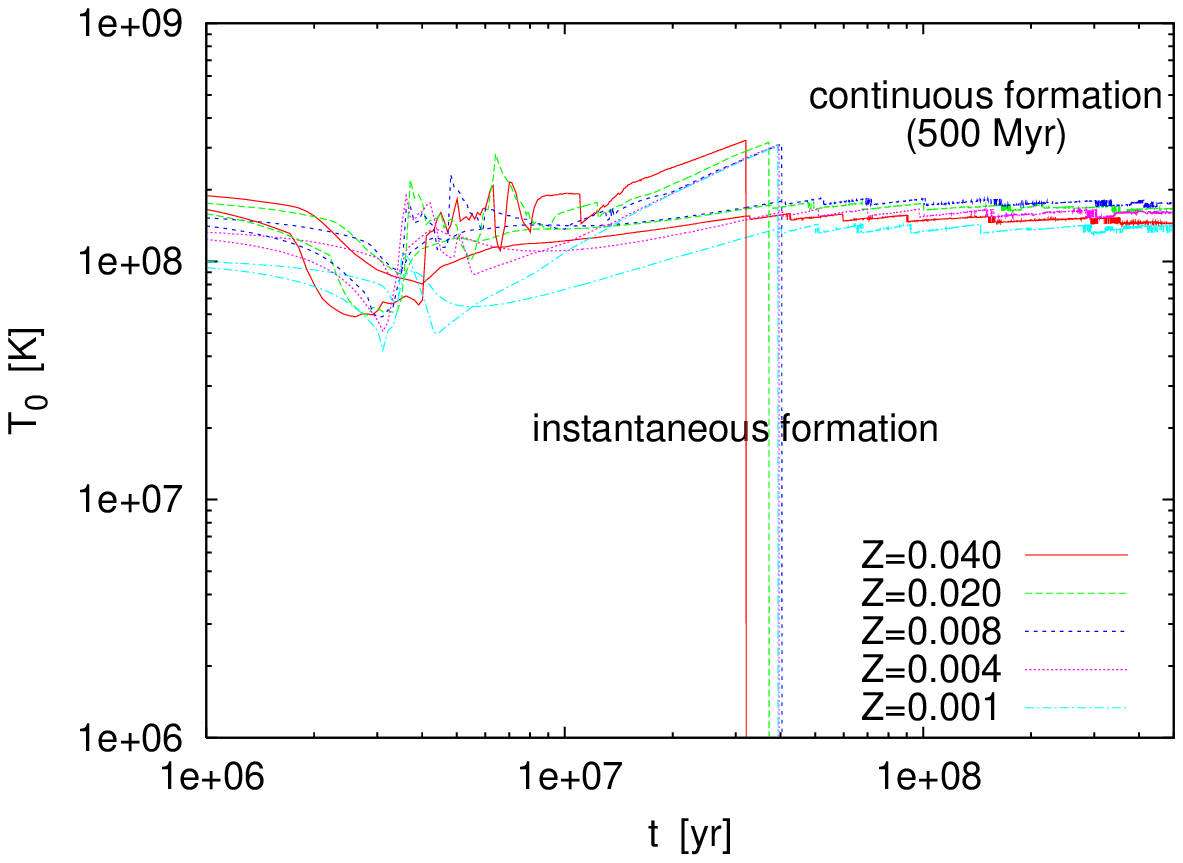}
  \caption{Evolution of the mass injection and heating parameters $\alpha$ and
  $T_0$ during galaxy formation for different metallicities $Z$, derived from
  Starburst99. The time-dependence of mass injection (\emph{top}) and heating
  parameter (\emph{bottom}) are shown for instantaneous formation of the galaxy
  and continuous formation during $\unit[500]{Myr}$. Heating by SNe Ia is not
  included. }
\end{figure}

\end{document}